# OptGM: An Optimized Gate Merging Method to Mitigate NBTI in Digital Circuits

**Amir M. Hajisadeghi*, Maryam Ghane, Hamid R. Zarandi**
Department of Computer Engineering, Amirkabir University of Technology (Tehran Polytechnic)

**Abstract:** This paper presents OptGM, an optimized gate merging method designed to mitigate negative bias temperature instability (NBTI) in digital circuits. First, the proposed approach effectively identifies NBTI-critical internal nodes—those with a signal probability exceeding a predefined threshold. Next, based on the proposed optimized algorithm, the sensitizer gate—which drives the critical node—and the sensitive gate, which is fed by it, are merged into a new complex gate. This complex gate preserves the original logic while eliminating NBTI-critical nodes. Finally, to evaluate the effectiveness of OptGM, we assess it on several combinational and sequential benchmark circuits. Simulation results demonstrate that, on average, the number of NBTI-critical transistors (i.e., PMOS transistors connected to critical nodes), NBTI-induced delay degradation, and the total transistor count are reduced by 89.3%, 24%, and 7%, respectively. Furthermore, OptGM enhances performance per cost (PPC) by 12.8% on average, with minimal area overhead.

## 1- Introduction

With continued technology scaling and thinner gate oxide layers, circuit aging has become a significant reliability challenge, particularly in nanometer-scale designs [1]. Negative bias temperature instability (NBTI) [2-4], positive bias temperature instability (PBTI) [5], hot carrier injection (HCI) [6, 7], and time-dependent dielectric breakdown (TDDB) [8] are the primary factors contributing to circuit aging and performance degradation. Among these, NBTI has gained increased attention as a critical reliability concern due to the aggressive downscaling of semiconductor technology nodes [1, 9]. This dominance stems from the exponential dependence of NBTI on oxide electric field combined with aggressive gate oxide scaling in sub-65nm nodes [1, 4, 9, 10]. Aging simulations for the 45*nm* technology node show that NBTI accounts for 68.2% of total delay degradation after 10 years under typical operating conditions, compared to PBTI: 17.8%, HCI: 10.3%, and TDDB: 3.7%. Therefore, while multiple aging mechanisms affect circuit reliability, NBTI is the main factor.

NBTI affects PMOS transistors when they are negatively biased, leading to degradation in electrical parameters such as increased threshold voltage, reduced transconductance, decreased linear and saturation current, lower channel mobility, and increased subthreshold swing [10]. Over time, these effects manifest as delay faults, potentially leading to system failure. NBTI follows a two-phase process: stress and recovery [1, 10]. During the stress phase, NBTI occurs when PMOS transistors are ON ($|V_{gs}| > |V_{th}|$), causing a significant increase in the absolute threshold voltage over time. Conversely, in the recovery phase, when PMOS transistors are OFF ($|V_{gs}| < |V_{th}|$), part of the degraded threshold voltage is recovered [1].

Several methods have been proposed to enhance the NBTI tolerance of digital circuits. Input vector control (IVC) methods manipulate input vectors to alter the zero probability of internal PMOS transistors [11-13]. However, their effectiveness is limited because internal nodes at deeper circuit levels are less controllable through input vectors. On the other hand, internal node control (INC) methods offer better controllability of internal nodes for NBTI mitigation [14-18]. The gate replacement method, a form of INC, replaces gates connected to NBTI-critical nodes to mitigate NBTI effects [17]. However, mostly due to the high design overhead, it can only be applied to a limited number of critical internal nodes. Another INC approach leverages sleep signals to control critical internal nodes by introducing transmission gates [15]. This method uses both PMOS and NMOS transistors, but the additional PMOS transistors remain susceptible to NBTI aging, leading to considerable delay, power, and area overheads.

Beyond digital logic, CMOS technology enables diverse applications spanning analog signal processing and intelligent computation. For instance, CMOS current squaring circuits have been successfully applied to analog signal processing tasks as conversion and frequency doubling [19]. Additionally, programmable implementations of complex nonlinear functions, such as fuzzy functions, have been realized efficiently in CMOS [20]. These examples underscore the technological breadth of CMOS circuits. However, regardless of the application domain—whether digital, analog, or mixed-signal—transistor aging phenomena like NBTI fundamentally threaten long-term reliability. While this work focuses on digital circuit mitigation, the gate-level optimization principles established here offer potential extensions to analog circuit reliability enhancement, particularly in aging-sensitive blocks such as operational amplifiers and comparators, where PMOS transistors experience prolonged stress phases.

* Corresponding author.
E-mail address: hajisadeghi@aut.ac.ir (Amir M. Hajisadeghi)

Based on the explanations, there is a clear need for an effective, low-overhead method to mitigate NBTI in digital circuits. Since NBTI-induced degradation strongly depends on the stress time of PMOS transistors, this paper presents a method that removes NBTI stress from transistors with long stress time. The proposed method first identifies NBTI-critical internal nodes by calculating their zero signal probability ($SP_0$). A node is classified as critical if its $SP_0$ exceeds a predefined threshold. Next, NBTI-critical nodes are removed by merging NBTI-sensitive gates and NBTI-sensitizer gates. The sensitive gate is driven by the critical node, while the sensitizer gate determines its value. Through this gate merging process, a new complex gate is formed with the same logic function, effectively eliminating NBTI-critical nodes.

The choice of $SP_0$ threshold significantly influences circuit performance and area overhead. A lower threshold classifies more nodes as critical, leading to increased circuit lifetime but also higher area overhead due to additional complex gates. Conversely, a higher threshold results in fewer critical nodes, offering minimal NBTI improvement. To balance performance and area overhead, this work introduces a performance per cost (PPC) metric, which optimizes the trade-off between circuit performance (delay) and area consumption. The method evaluates $SP_0$ threshold values from {0.5, 0.65, 0.75, 0.85, 0.95}, selecting the one that maximizes PPC. Using this optimization, the average PPC across circuits improves by 12.8% compared to a fixed threshold value of 0.5.

The proposed method was evaluated on a subset of circuits from different benchmark suites. Results show that it reduces NBTI-critical transistors (i.e., PMOS transistors connected to critical nodes) by 89.3%, lowers the total transistor count by 7.03% on average, and mitigates NBTI-induced delay degradation by 24.03% over 10 years of operation. Despite these improvements, it incurs only a 1.96% area overhead, demonstrating its efficiency and practicality for NBTI mitigation. The key contributions of this paper are as follows:

- Proposing OptGM, an optimized gate merging method for NBTI mitigation in digital circuits.
- Introducing the PPC metric to select the optimal $SP_0$ threshold by balancing NBTI mitigation with delay and area overhead.
- Designing an algorithmic framework that identifies NBTI-critical internal nodes and merges sensitizer and sensitive gates into complex gates.
- Conducting a comprehensive evaluation that demonstrates the high efficiency of the method in reducing NBTI degradation with negligible area overhead.

The remainder of this paper is structured as follows. Section 2 provides background on NBTI phenomena and NBTI models. Section 3 reviews related work. Section 4 describes the proposed NBTI mitigation method in detail. Section 5 presents evaluation results on benchmark circuits. Finally, Section 6 concludes the paper.

## 2- Background

NBTI causes degradation in the key electrical parameters of stressed devices, such as threshold voltage, transconductance, linear and saturation current, channel mobility, and subthreshold swing [10]. This section first describes the NBTI phenomenon and then discusses the main factors that intensify it.

### 2-1- NBTI Phenomenon

A widely used model to explain NBTI is the Reaction-Diffusion (R-D) model [10, 21, 22], which highlights the central role of hole-trapping mechanisms in NBTI behavior. According to this model, NBTI occurs when a PMOS transistor is turned ON ($|V_{gs}|>|V_{th}|$) and a vertical electric field is present across the gate oxide. During the stress phase, when a negative bias is applied to the gate of a PMOS transistor, relatively weak Si–H bonds at the silicon-oxide interface are broken due to the electric field. This causes hydrogen atoms to diffuse into the dielectric, forming positive interface traps and molecular hydrogen [10, 21]. The positive charge accumulation shifts the threshold voltage in the positive direction, which contributes to circuit aging. As a result, the absolute value of the threshold voltage increases significantly over time [1, 10].

A particular characteristic of NBTI is its partial recoverability. In the recovery phase, when the PMOS transistor is OFF ($|V_{gs}|<|V_{th}|$), hydrogen diffuses back toward the interface and anneals the broken silicon bonds, allowing some of the Si–H bonds to be reformed. This reduces the number of interface traps and leads to partial recovery of the threshold voltage [10, 21].

Fig. 1 illustrates four major factors that influence NBTI: 1) stress time, 2) signal probability, 3) temperature, and 4) vertical electric field. Fig. 1(a) shows the threshold voltage variation over time during the stress and recovery phases for 90*nm* PMOS transistors [1, 10, 21]. In the stress phase, the threshold voltage increases, aging the PMOS transistors and degrading circuit performance. In the recovery phase, the threshold voltage partially recovers [1, 10, 21]. Stress time affects threshold voltage with time radically ($\Delta V_{th} \propto t^n$, $n \approx 0.25$). Moreover, zero signal probability ($SP_0$) increases the absolute value of threshold voltage dramatically ($\Delta V_{th} \propto (SP_0)^n$). Temperature (Fig. 1(b)) and vertical electric field (Fig. 1(c)) also have an impact on NBTI. These parameters increase the absolute value of threshold voltage exponentially ($\Delta V_{th} \propto e^T$, $\Delta V_{th} \propto e^{Eox}$) [1, 10, 21].

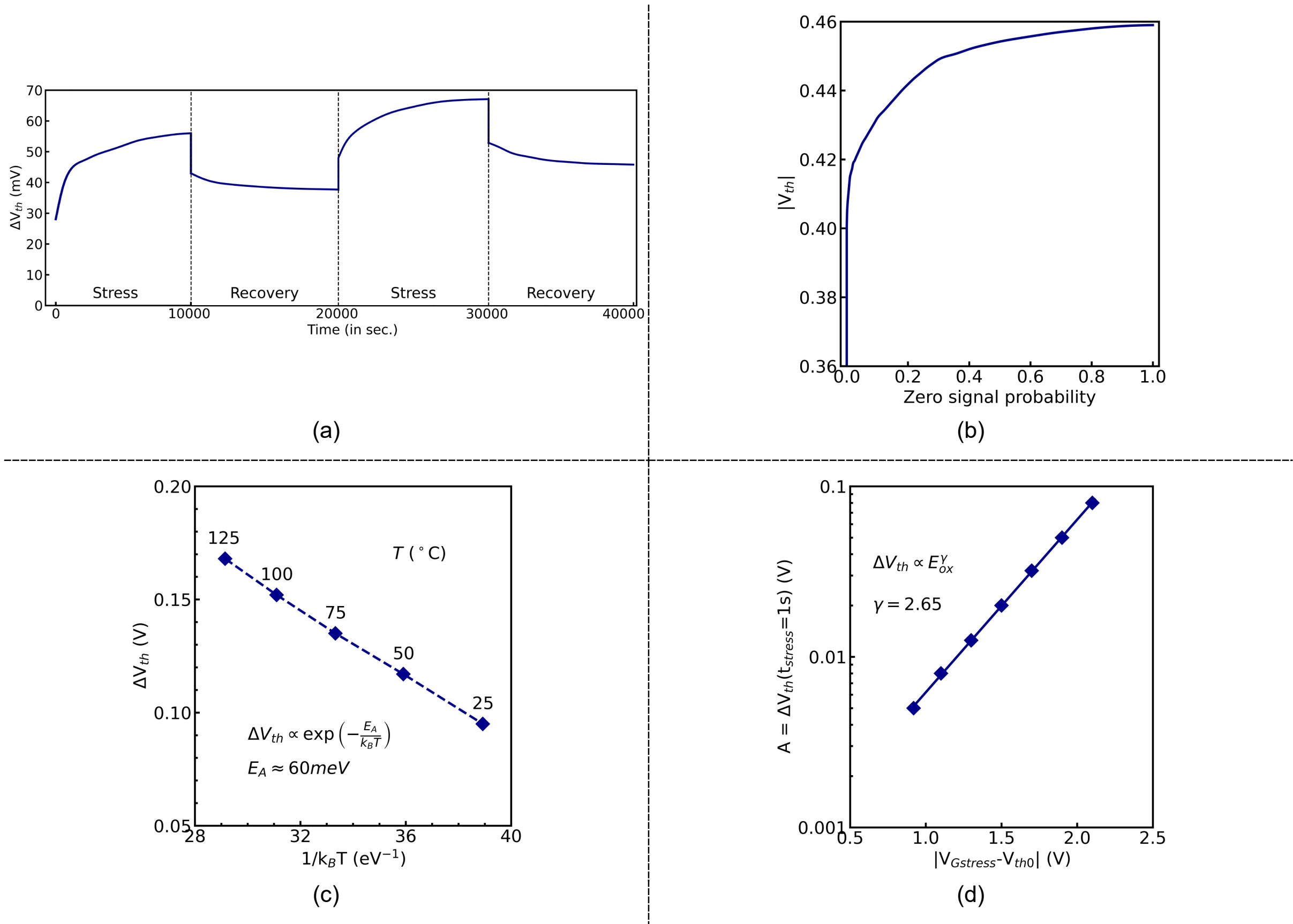

Fig. 1. Four major factors in the occurrence of NBTI: a) stress time, b) zero signal probability, c) temperature, and d) vertical electric field.

### 2-1- NBTI Model

The variation in threshold voltage ($\Delta V_{th}$) during the stress and recovery phases can be modeled by the following equations [1]:

$$\Delta V_{th}(t) = \left( K_v (t\text{-}t_0)^{0.5} + \sqrt[2n]{\Delta V_{th}(t_0)} \right)^{2n} \quad (1)$$

$$\Delta V_{th}(t) = \Delta V_{th}(t_1) \left( 1\text{-} \frac{2\varepsilon_1 . t_e + \sqrt{\varepsilon_2 . C.(t\text{-}t_1)}}{(1+\delta).t_{ox} + \sqrt{C.t}} \right) \quad (2).$$

Here, $t_0$ and $t_1$ denote the start times of the stress and recovery phases, respectively, and $t$ is the elapsed stress time. $K_v$ depends on temperature and electric field; $\varepsilon_1$, $\varepsilon_2$, and $C$ are temperature-dependent constants; $t_{ox}$ and $t_e$ denote the oxide thickness and effective oxide thickness, respectively; $\delta$ is a fitting parameter less than 1.

A closed-form expression for the overall $\Delta V_{th}$ after accounting for both stress and recovery phases can be given as [1]:

$$\Delta V_{th}(t)=\left(\frac{\sqrt{K_v^2.\alpha.T_{clk}}}{1-\beta_t^{\frac{1}{2n}}}\right)^{2n} \tag{3}$$

$$\beta_t=1-\frac{2\varepsilon_1.t_e+\sqrt{\varepsilon_2.C.(1-\alpha).T_{clk}}}{2t_{ox}+\sqrt{C.t}} \tag{4}$$

where $\alpha$ denotes the stress probability, $T_{clk}$ is the clock period, and $n$ typically varies between 0.1 and 0.25, depending on the technology. Thus, stress probability ($\alpha$) plays a critical role in determining the severity of NBTI and is a key parameter in designing NBTI mitigation strategies for digital circuits [2, 21].

## 3- Related Work

Several methods have been proposed to enhance the tolerance of digital circuits against NBTI. As shown in Fig. 2, these methods are generally categorized into two main groups: 1) compensation methods, and 2) mitigation methods. Compensation methods typically utilize techniques such as transistor/gate sizing [23, 24] threshold voltage ($V_{th}$) tuning [22, 25], and guard banding [26, 27] to counteract the destructive effects of NBTI. These approaches aim to extend circuit lifetime by minimizing the impact of NBTI on device parameters. On the other hand, mitigation methods can be further divided into design-time methods and adaptive methods. Design-time techniques include NBTI-aware logic synthesis [22, 28, 29], input vector control (IVC) [11-13], and internal node control (INC) methods [14-18].

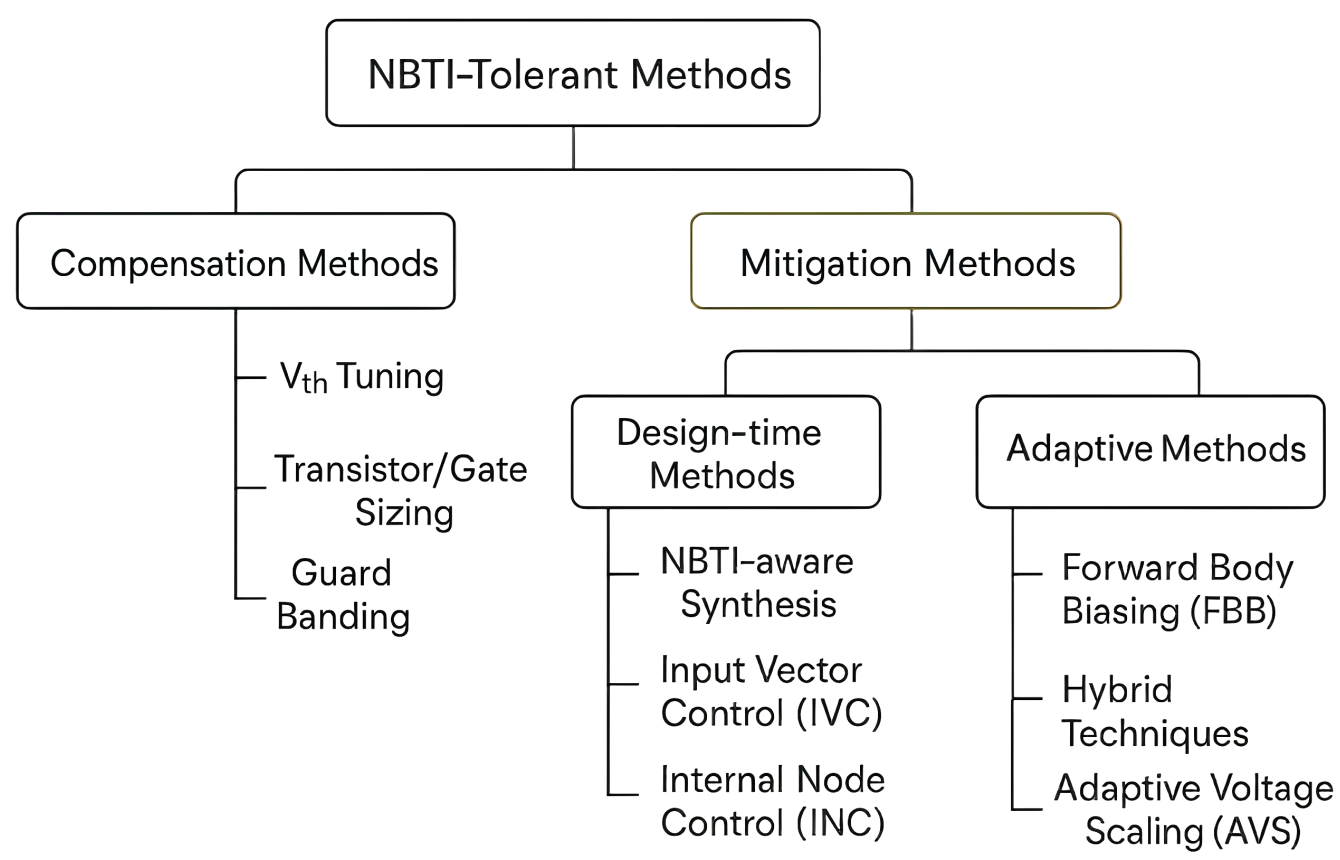

Fig. 2. Classification of NBTI-tolerant methods.

The method proposed in [30] introduces an NBTI-aware digital low-dropout regulator with adaptive gain scaling control, achieving a 33% improvement in mitigating NBTI effects. Several IVC-based approaches [11-13] aim to reduce NBTI degradation by adjusting input vectors and performing transistor reordering. These methods modify the zero signal probability of internal nodes to reduce stress on PMOS transistors. However, their effectiveness is limited in large circuits because internal nodes at deeper levels are less controllable via external input vectors, and consequently, NBTI mitigation applies to only a small subset of transistors.

INC-based approaches address this limitation by directly controlling internal nodes to suppress NBTI effects. Some INC methods achieve this by inserting sleep transistors into gates or internal paths [15, 31]. For instance, the method in [15, 31] proposes adding a sleep signal that controls internal nodes during standby mode. Bild et al. [31] assume that the system can generate sleep signals to load new external inputs and activate pre-selected internal signals for internal node control. This method modifies pull-up and pull-down networks by inserting NMOS and PMOS transistors. Although it reduces NBTI, it also increases leakage power and necessitates structural changes to all gates in the library, requiring custom gate redesign and replacement.

Wang et al. [17] propose a gate replacement technique using existing library gates to insert a sleep signal. However, this method can only be applied to a limited number of NBTI-critical internal nodes due to library constraints. Moreover, it does not account for timing overheads introduced by the gate replacement, which may adversely affect the critical path delay. Another INC-based method [15] proposes controlling internal nodes by inserting sleep signals using two PMOS and one NMOS transistor. While this reduces NBTI, the added PMOS transistors are themselves prone to NBTI-induced degradation, and the method imposes substantial delay and power overhead.

The method presented in [32] uses an adaptive body biasing technique and achieves significant NBTI reduction, decreasing total yield loss by 97.1% after 10 years. However, it increases both dynamic power and leakage power during the first five years. Other adaptive methods [33, 34] use forward body biasing (FBB) to reduce threshold voltage and improve performance. However, this exponentially increases subthreshold leakage. Dynamic voltage and frequency scaling (DVFS) techniques [35] are also utilized to manage threshold voltage dynamically. Consequently, hybrid techniques that combine FBB and adaptive voltage scaling [36] have been proposed to achieve better trade-offs between power consumption and NBTI mitigation.

In this paper, under the INC category, we propose a gate-merging approach that aims to be more effective and efficient than the existing INC methods, which often incur high overhead. Method [14] previously introduced a gate-merging technique to mitigate NBTI in digital circuits. While that approach effectively reduces NBTI, it relies on a static and unoptimized selection of NBTI-critical internal nodes for merging. In contrast, the present study advances the concept by 1) introducing a performance-per-cost metric that formulates threshold selection as an optimization problem rather than a fixed assumption, 2) providing a systematic algorithmic framework for scalable gate merging across both critical and non-critical paths, and 3) conducting a broader evaluation across different benchmark circuits under long-term aging evaluation. These extensions transform the earlier concept into a practical and reproducible flow that ensures, for each circuit, an optimized trade-off between NBTI mitigation and design overhead.

## 4- The Proposed Method: Optimized Gate Merging

This section introduces OptGM, an optimized gate merging method to mitigate NBTI-induced degradation in digital circuits by eliminating NBTI-critical internal nodes. The proposed method comprises two main phases. First, it identifies NBTI-critical PMOS transistors—those with high-stress probability—by estimating the zero signal probability ($SP_0$) of internal nodes. A transistor is considered critical if it remains ON for a significant portion of time, which corresponds to an internal node with an $SP_0$ above a predefined threshold.

Once the NBTI-critical internal nodes are identified, OptGM removes them by merging their driving and driven gates into a new complex gate that implements the equivalent logic. This process eliminates the critical node and its associated PMOS transistor, thereby reducing NBTI-induced degradation.

The threshold for determining criticality is circuit-specific and selected using the performance per cost metric, which balances delay reduction against area overhead. A higher threshold removes more transistors but increases complexity; a lower one reduces overhead but limits NBTI improvement.

### 4-1- Fundamental Concepts and Definitions

Before detailing the proposed optimized gate merging algorithm, the key terms and assumptions are defined.

**Signal Probability:** The probability that a node holds logic '0' or '1', denoted as $SP_0$ and $SP_1$, respectively. For a node $n$, $SP_0(n) = 1 - SP_1(n)$. $SP_0$ is used to estimate the stress probability of PMOS transistors connected to internal nodes. A higher $SP_0$ indicates the transistor is ON longer, contributing more to NBTI. $SP_0$ values for internal nodes are derived from the signal probabilities of primary logic gates, as shown in Table 1.

Table 1. $SP_1$ calculation of primary gates

| Primary gate | Inputs $SP_1$ | Output $SP_1$ calculation |
|---|---|---|
| *NOT* | $S_1$ | $1-S_1$ |
| *AND* | $S_1$, $S_2$ | $S_1 \times S_2$ |
| *OR* | $S_1$, $S_2$ | $S_1+S_2-(S_1 \times S_2)$ |

**NBTI-Critical Node:** An internal node is considered NBTI-critical if its $SP_0$ exceeds a predefined threshold. This threshold reflects the acceptable upper limit of transistor stress in the circuit. Critical nodes are targeted for elimination through gate merging.

**Threshold Value:** A tunable parameter that defines the $SP_0$ boundary for node criticality. Lower values mark more nodes as critical—enhancing mitigation but potentially increasing the area. The optimal threshold is selected using:

$$\text{Performance Per Cost (PPC)} = \frac{\frac{1}{\text{Delay}}}{\text{Area}} = \frac{1}{\text{Delay} \times \text{Area}} \quad (5).$$

In this paper, NBTI-induced degradation is calculated for different threshold values. The threshold value that the PPC of the modified circuit has a maximum value is considered the best threshold value for that circuit. It should be noted that in many high-performance digital systems—such as datapath components in processors, real-time control logic in aerospace and industrial applications, and hardware accelerators for signal processing—delay and silicon area are the primary design constraints. While delay directly impacts system responsiveness and throughput, minimizing area reduces cost, enables higher integration, and satisfies form-factor constraints in embedded and mission-critical systems. Accordingly, the proposed PPC metric emphasizes these factors to align with practical design priorities in such NBTI-aware scenarios.

**Sensitive Gate:** A gate is sensitive if one of its inputs is an NBTI-critical internal node. For instance, in Fig. 3(a), the $SP_0$ of node $N_5$ is 0.75, and if the threshold is set to 0.75, then $N_5$ is an NBTI-critical node. Therefore, gate $G_3$ is a sensitive gate.

**Sensitizer Gate:** A gate is a sensitizer if its output is an NBTI-critical node, meaning it directly drives a high-$SP_0$ node and hence contributes to the stress of the connected PMOS transistor. For example, in Fig. 3(a), gate $G_1$ is the sensitizer of node $N_5$.

**Complex Gate:** A gate formed by merging sensitizer and sensitive gates. It preserves the logic function and eliminates the internal node. Fig. 3(b) illustrates a complex gate generated by merging AND and NAND gates.

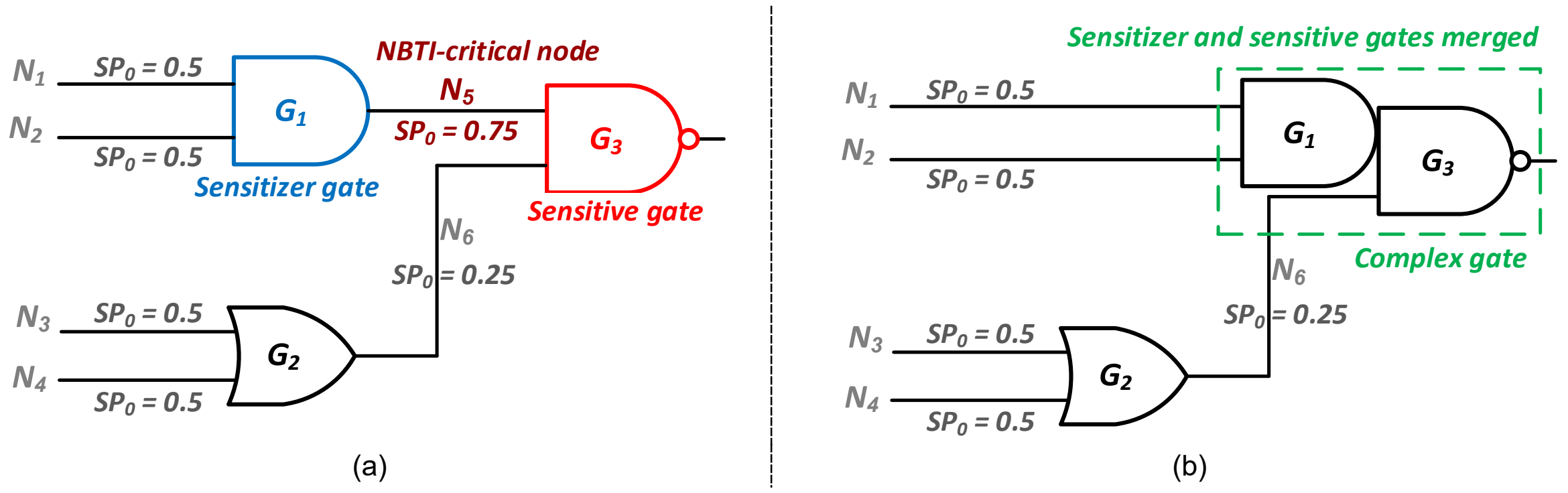

Fig. 3. An illustration of a) NBTI-critical node, sensitive and sensitizer gates, and b) merging of AND and NAND gates to generate a complex gate with logic function $F = ((N_1.N_2).(N_3+N_4))'$.

### 4-2- Optimal Gate Merging Algorithm

In order to reduce the NBTI-induced degradation, an optimal gate merging algorithm is proposed. It follows three main steps for each circuit: 1) NBTI criticality analysis, 2) gate merging, and 3) optimal threshold determination. Algorithm 1 coordinates these steps by invoking the *NBTI_Criticality_Analysis* (Algorithm 2) and *Gate_Merging* (Algorithm 3) functions for each candidate threshold in a predefined set. Then, the threshold value that maximizes the PPC is selected as the optimal threshold value and used for the final application.

**Algorithm 1:** Optimal Gate Merging

***Input:*** *Netlist of the circuit*
***Output:*** *Optimized NBTI-tolerant circuit*

```
1   Threshold Set = {0.5, 0.65, 0.75, 0.85, 0.95};
2   OGM_Circuit_Set = {};
3   foreach T in Threshold Set do
4   |  (NBTI_Critical_Nodes, Sensitizer_Gates, Sensitive_Gates) = NBTI_Criticality_Analysis(Circuit, T);
5   |  GM_Circuit = Gate_Merging (Circuit, NBTI_Critical_Nodes, Sensitizer_Gates, Sensitive_Gates);
6   |  GM_PPC = Calculate PPC for GM_Circuit;
7   |  Insert (GM_Circuit, GM_PPC) to OGM_Circuit_Set;
8   end foreach
9   OGM_Circuit = Circuit with maximum PPC from OGM_Circuit_Set;
10  return (OGM_Circuit);
```

**Algorithm 2:** NBTI Criticality Analysis of Circuit (***NBTI_Criticality_Analysis***)

***Input:*** *Netlist of the circuit* $G = [g_1, g_2, \ldots, g_n]$*, and threshold value (T)*
***Output:*** *NBTI-critical nodes, sensitizer, and sensitive gates*

```
G_sorted = topological_order_sort(G);
L = depth of netlist;
N = [n_1, n_2, …, n_m];  // Internal nodes of circuit.
// Calculate SP0 of internal nodes of circuit.
SP_0(primary inputs) = 0.5;
for i=1 to L do
│ foreach g in G_sorted(i) do
│ │ Calculate SP_0 of the output node of the gate g;
│ end foreach
end for
// Determining NBTI-critical nodes.
foreach n in N do
│ if SP_0(n) ≥ T then
│ │ Node n is the NBTI-critical node;
│ end if
end foreach
// Determining sensitive and sensitizer gates.
foreach g in G do
│ if Output(g) is NBTI-critical node then
│ │ Gate g is sensitizer gate;
│ end if
│ if Input(g) is NBTI-critical node then
│ │ Gate g is sensitive gate;
│ end if
end foreach
return (NBTI_Critical_Nodes, Sensitizer_Gates, Sensitive_Gates);
```

**Algorithm 3:** Merging Sensitizer and Sensitive Gates (***Gate_Merging***)

***Input:*** *Netlist of the circuit, NBTI-critical nodes, sensitizer, and sensitive gates*
***Output:*** *NBTI-tolerant circuit*

```
F = [f_1, f_2, …, f_N];   // Number of fan-out branches of node n.
Critical_Path = Find critical path of circuit;
Sorted_NBTI-critical_nodes_CP = Sort NBTI-critical nodes in critical path of circuit in descent order
    based on SP_0 value;
Non_Critical_Path = Find non-critical paths of circuit;
Sorted_NBTI-critical_nodes_NonCP = Sort NBTI-critical nodes in critical path of circuit in descent
    order based on SP_0 value;
foreach S in [Sorted_NBTI_critical_nodes_CP, Sorted_NBTI_critical_nodes_NonCP] do
│ foreach n in S do
│ │ if sensitizer AND sensitive gates of node n are not complex gates then
│ │ │ if F(n) > 1 then
│ │ │ │ Create F(n)-1 copy of sensitizer gate of node n;
│ │ │ end if
│ │ end if
│ │ Convert into one complex gate (sensitizer gate of node n, sensitive gate of node n);
│ end foreach
end foreach
return (GM_Circuit); // Gates merged circuit.
```

Algorithm 2 describes the NBTI criticality analysis step. It begins with levelizing the circuit gates based on topological order. The $SP_0$ of primary inputs is assumed to be 0.5 (though it may vary based on real input activity). $SP_0$ values of all internal nodes are computed recursively using the gate-level structure and Table 1. Nodes exceeding the threshold are marked as NBTI-critical, and their sensitizer and sensitive gates are recorded.

Algorithm 3 describes the gate merging step. First, the NBTI-critical internal nodes of the circuit are identified separately for the critical path and the non-critical paths. Then, within each group, the nodes are sorted in descending order based on their $SP_0$ values. This prioritization ensures that the nodes with the highest stress probability are handled first. Subsequently, the algorithm iterates over the sorted lists, giving priority to the critical path nodes, followed by the non-critical path nodes. For each selected NBTI-critical node, if both the associated sensitive and sensitizer gates are not already complex, they are merged into a new complex gate that preserves the original logic functionality. This merging process involves generating the Boolean expressions of the gates and combining them according to the complex gate generation rules [37], [38]. It should be noted that if a sensitizer gate has $n$ fan-outs, $n-1$ additional copies are created for merging. Also, merging is prioritized on the critical path to maximize delay benefits.

The proposed algorithms were evaluated on a wide range of benchmark circuits; nevertheless, formal analysis showing the correctness and completeness of the algorithms has been presented in the Appendix. The overall timing complexity of Algorithm 1 is dominated by the sorting procedure in *Gate_Merging* function (Algorithm 3), which is $O(N \times \log N)$, where $N$ is the number of NBTI-critical nodes in the netlist. However, because the proposed optimal gate merging algorithm is executed during the implementation phase for each circuit, its time complexity would not be challenging. With respect to scalability, the overall timing complexity of OptGM could be acceptable for design-time optimization but potentially heavy for highly iterative flows. To address this, a hierarchical decomposition strategy, which means partitioning the design into functional blocks and executing OptGM per block in parallel, can be employed.

It is worth noting that the effectiveness of the proposed method becomes more prominent as the number of primary gates in the synthesized circuit increases. When a circuit is synthesized with a larger proportion of complex gates instead of primary gates, the method remains applicable and continues to mitigate NBTI-induced degradation. In such cases, the area overhead may increase due to the merging of complex gates with high-fanout gates that are not as optimized as those available in the library technology file. Nevertheless, the approach still preserves its reliability benefits. Furthermore, in scenarios dominated by complex gates, the proposed method can be extended to the transistor level by designing NBTI-aware complex standard cells with OptGM incorporated directly into their transistor design, offering an additional design opportunity.

Regarding the integration of the proposed method with ASIC design flow, this method can be incorporated with synthesis algorithms as an NBTI-aware synthesis process to reach a synthesized circuit with high robustness against NBTI phenomena. Alternatively, it can be applied after the synthesis process and before the placement process of the design flow. Fig. 4 illustrates these two approaches overlaid on the standard ASIC design flow.

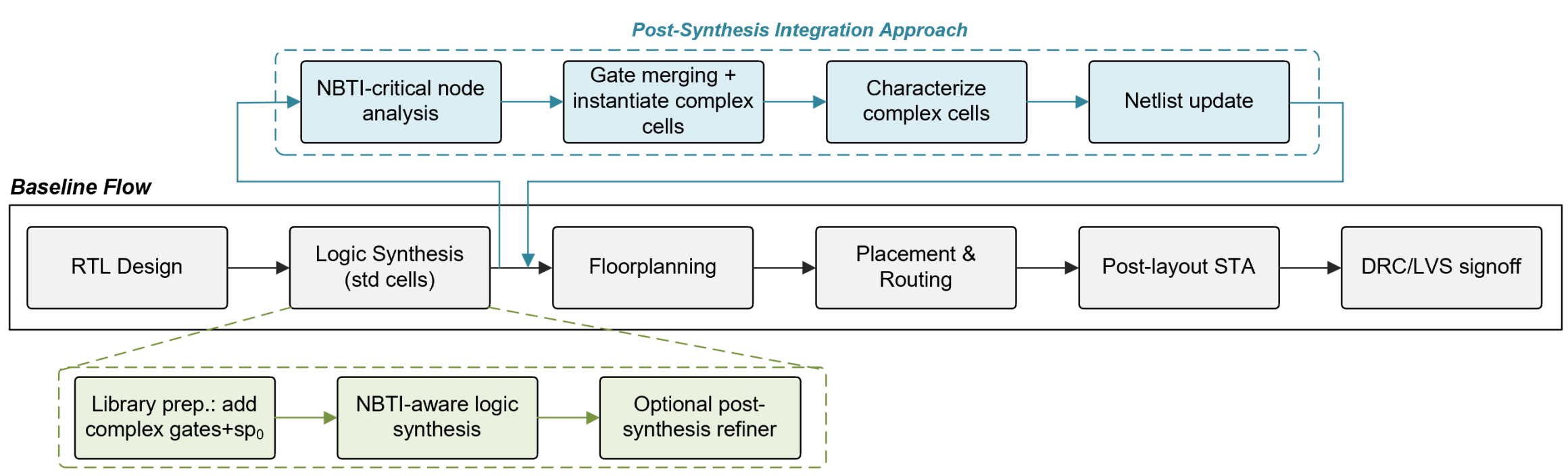

Fig. 4. An illustration of a composite ASIC design flow diagram including: OptGM post-synthesis integration (in blue), and NBTI-aware synthesis (in green) approaches.

In the post-synthesis integration approach, OptGM is applied to the synthesized gate-level netlist: NBTI-critical nodes are identified, selected merges are instantiated as complex gates, those gates are characterized, the netlist is updated and equivalence-checked, and an incremental synthesis run completes timing closure. This path requires custom cell characterization and occasional timing re-closure, but is non-intrusive (no modification of existing implementation tools) and is well-suited to academic studies and retrofitting existing designs.

For production deployment, the NBTI-aware synthesis approach is recommended: a one-time library preparation adds a modest set of pre-characterized complex primitives and $SP_0$ annotations, and the synthesis engine is extended (via a lightweight plugin) to include an NBTI term in the cost function (Cost = $\alpha$×Delay + $\beta$×Area + $\gamma$×$NBTI_{Factor}$); the engine then produces NBTI-optimized netlists directly, avoiding post-hoc replacement and reducing timing re-closure risk.

However, with respect to electronic design automation (EDA) integration, two primary challenges arise. 1) Custom cell generation: OptGM may introduce complex cells that are not available in foundry libraries. Depending on the chosen integration strategy, these cells must either be characterized and the place-and-route flow re-executed, or a standard-cell library containing pre-designed complex primitives—with NBTI-aware cost terms embedded in the technology-mapping stage—must be employed. This approach ensures timing closure during synthesis but requires tool-level support. 2) Timing-model accuracy: Because complex gates exhibit non-linear delay behavior, traditional static timing analysis (STA) can underestimate their delay. To address this issue, a non-linear delay model (NLDM) is required for the commonly merged gates, and its accuracy can be validated against HSPICE simulations.

Fan-out duplication is the main source of area overhead in OptGM, and several mitigation strategies can reduce this overhead with minimal impact on effectiveness. First, a selective-duplication policy with timing priority processes critical-path nodes before non-critical ones. For nodes with fan-out greater than one, it identifies timing-critical branches and creates duplicates only for those branches while leaving non-critical branches connected to the original sensitizer; this approach suits designs with tight area budgets that can tolerate a small loss in NBTI mitigation. Second, shared complex-gate designs replace duplicated sensitizers by multi-output complex cells; although this reduces transistor count, it increases routing and placement complexity and requires custom cell support, so it is most appropriate in full-custom ASIC flows where library extension is acceptable. Third, for high fan-out, a buffer-tree alternative can be used instead of duplication: buffers generally incur lower area than full duplication and introduce less delay than a duplicated gate cascade, but they do not remove the original critical node and therefore yield reduced NBTI mitigation. In practice, a hybrid strategy that combines selective duplication, selective buffering, and selective use of shared complex cells is recommended; the hybrid approach attained minimal area overhead while sacrificing a negligible amount of NBTI mitigation effectiveness, providing an attractive trade-off for area-constrained designs.

In the remainder of this section, two illustrative examples are provided to further clarify the proposed gate merging method. In the first example, Fig. 5(a) shows a 3-input transistor-level NOR circuit that implements the logic function $((A + B)' + (C + D)')'$. Assuming the $SP_0$ of all inputs is 0.5, the resulting $SP_0$ values of nodes $m$ and $n$ are each 0.75. This means that, statistically, the PMOS transistors connected to nodes $m$ and $n$ are ON 75% of the time. If the $SP_0$ threshold is set to 0.75, nodes $m$ and $n$ are classified as NBTI-critical internal nodes. Consequently, gates $G_1$ and $G_2$ are identified as sensitizer gates, and $G_3$ is the corresponding sensitive gate.

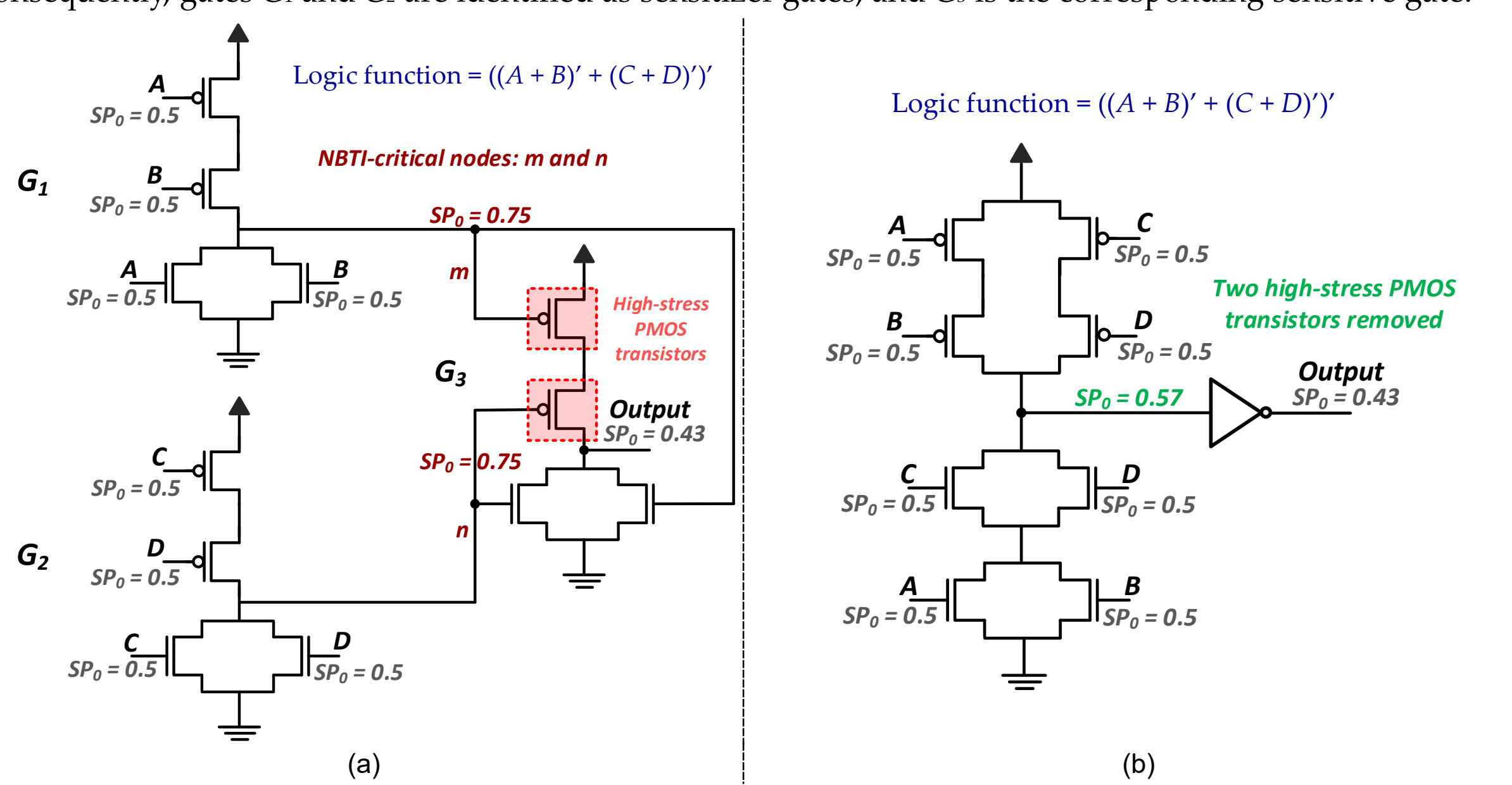

Fig. 5. An example of converting sensitizer and sensitive gates into a complex gate and removing internal NBTI-critical nodes; a) 3-input NOR circuit, and b) its equivalent NBTI-tolerant circuit.

According to NOR logic, the Boolean expressions for $G_1$ and $G_2$ are $G_1 = (A + B)'$ and $G_2 = (C + D)'$, respectively. The output of $G_3$, therefore, evaluates the expression $(G_1 + G_2)'$ which simplifies to $((A + B)' + (C + D)')'$. Based on the complex gate generation rules [37, 38], this logic expression is converted into a single complex gate. By merging the sensitizer and sensitive gates, the internal NBTI-critical nodes (*m* and *n*) are removed. The final modified circuit is shown in Fig. 5(b), where the transformation reduces the number of stressed PMOS transistors and hence mitigates NBTI-induced degradation.

A second example is presented in Fig. 6(a), which depicts a gate-level combinational circuit. Based on the $SP_0$ values shown, and assuming an optimal threshold value of 0.75, internal nodes $N_7$ and $N_8$ are identified as NBTI-critical nodes. Accordingly, gates $G_1$ and $G_2$ are classified as sensitizer gates, while gates $G_3$, $G_4$, and $G_5$ are identified as sensitive gates. The PMOS transistors in $G_3$, $G_4$, and $G_5$ are therefore subject to significant NBTI stress. In this circuit, the critical path is $G_0 \rightarrow G_2 \rightarrow G_4 \rightarrow G_7 \rightarrow G_8$. According to Algorithm 3, because sensitizer gates $G_1$ and $G_2$ each have two fan-out branches, duplicate versions of these gates ($G_1'$ and $G_2'$) are instantiated to enable individual merging with their respective sensitive gates (as shown in Fig. 6(b)). Also, since gate G0 also exhibits a fan-out of two in this scenario, a duplicate instance of $G_0$ (denoted $G_0'$) is likewise appended to the netlist to enable independent merging with its corresponding sensitive branch.

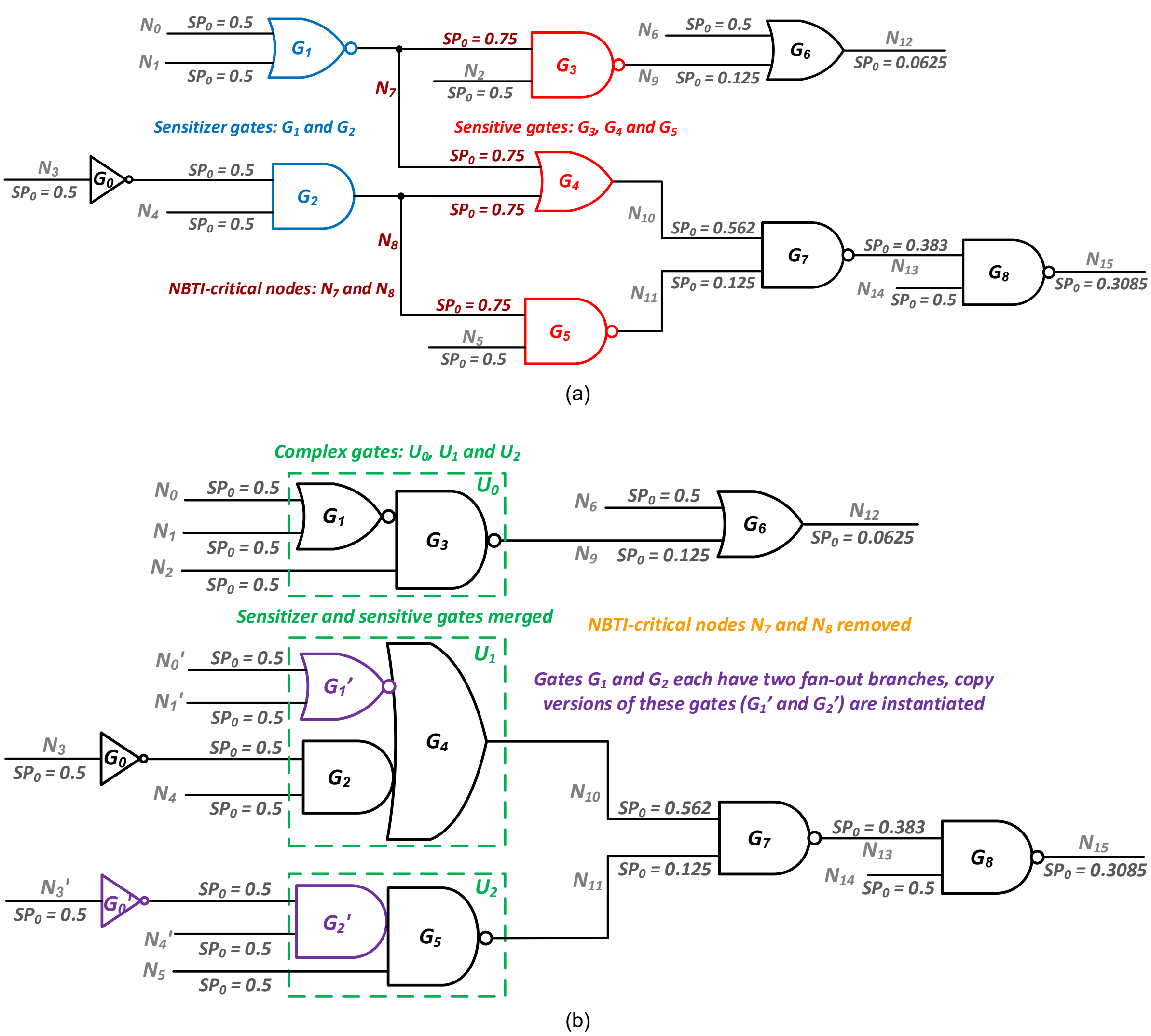

Fig. 6. An example of applying the proposed gate merging method to a combinational circuit; $SP_0$ values of the circuit node: a) before gate merging, and b) after gate merging.

In the critical path, gates $G_1'$ and $G_2$ are merged with sensitive gate $G_4$ to form complex gate $U_1$, while $G_2'$ and $G_5$ are merged into complex gate $U_2$. These complex gates are then inserted into the circuit, replacing the original ones. Subsequently, in the non-critical paths, gates $G_1$ and $G_3$ are merged to form complex gate $U_0$, which also replaces the original components. Through this transformation process, all identified NBTI-critical internal nodes and their associated PMOS transistors are effectively eliminated. As a result, the overall NBTI stress within the circuit is reduced, and the method successfully mitigates NBTI-induced degradation.

To clearly illustrate the Boolean expression merging methodology, consider the function $U_1 = (A+B)' + (C.D)$ in Fig. 6(b) as an example. Applying De Morgan to the complemented sum gives $(A+B)' = A'.B'$, so $U_1 = (A'.B') + (C.D)$. This sum-of-products form matches an OAI22 pattern when the second pair of inputs is pre-inverted. Hence, $U_1 = ((A+B).(C'+D'))'$ obtains by a OAI22 cell as OAI22(A, B, C′, D′). This procedure can be generally employed for any function by rewriting the expression with De Morgan/Boolean identities to expose an OAI/AOI or multi-input canonical form, and then mapping it to the single complex primitive that matches that form. If the considered complex gate is not available in the target cell library, equivalent implementations are possible, but the direct mapping is preferred when the primitive exists, as it minimizes area, critical-path depth, and switching power.

## 5- Simulation Result

The effectiveness of the proposed method has been evaluated on a subset of circuits from the ISCAS'85, ISCAS'89, and ITC'99 benchmark suites. The algorithms were implemented in the C programming language. Each benchmark circuit was simulated using 10,000 random input vectors. Transistor-level simulations incorporating NBTI modeling were conducted using the MOS reliability analysis (MOSRA) model in Synopsys HSPICE, based on the 45*nm* planar predictive technology model (PTM) [39, 40]. The supply voltage was set to 1V, and all circuits were subjected to NBTI stress for 10 years. Area comparisons were calculated based on the sum of transistor widths, which is an accepted metric for planar CMOS technologies [41].

In this work, the focus was placed on the dynamic NBTI effects, which were modeled using the MOSRA model. This approach incorporates both stress and recovery phases, allowing for accurate simulation of NBTI degradation under circuit operation with wide varying input vectors. It provides a more realistic assessment of NBTI impact compared to static conditions, which overestimate degradation by ignoring recovery effects.

It should be mentioned that OptGM targets combinational logic paths between flip-flops, focusing on NBTI mitigation in logic gates. Nevertheless, this approach remains effective and provides indirect sequential benefits. Even without modifying flip-flops, OptGM contributes to sequential reliability in two key ways: 1) it reduces setup and hold violations by mitigating combinational path delay degradation, thereby improving timing margins at flip-flop inputs; and 2) it alleviates clock tree stress, as gate merging lowers toggle activity on internal nodes that drive clock-gating logic, reducing the load and aging stress on clock buffers. Moreover, while OptGM currently focuses on combinational logic, employing the gate merging method beside sequential elements represents a promising direction. Preliminary investigation suggests applying controlled fan-out duplication to feedback paths in flip-flops could reduce internal PMOS stress, though careful clock domain analysis would be required. The hybrid approach, combining OptGM with emerging double-node-upset-recovery latches [42] and SEU-tolerant designs [43], could achieve holistic aging resilience with minimal overhead.

The proposed gate merging method removes NBTI-critical internal nodes, thereby eliminating transistors with high $SP_0$ values from the modified circuits. However, to ensure an effective reduction in NBTI-induced delay degradation, it is necessary to first determine the optimal $SP_0$ threshold value for each circuit. Fig. 7 illustrates the normalized PPC for various threshold values applied to the *s298* circuit. The PPC values have been normalized with respect to the PPC of the unmodified (base) circuit.

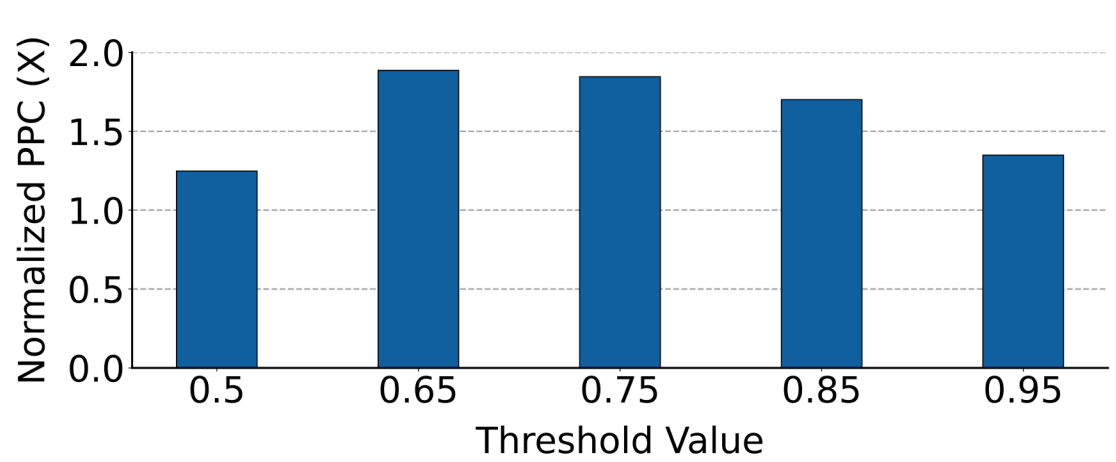

Fig. 7. Normalized PPC to base circuit for various threshold values in the gate merging method applied to the *s298* circuit.

In this example, the threshold value 0.65 yields the maximum PPC and is therefore selected as the optimal threshold. It is notable that for all tested thresholds, the PPC of the modified circuit exceeds that of the base circuit, indicating that the delay reduction outweighs the area overhead introduced by gate merging.

Additionally, Fig. 8 presents the $SP_0$ distribution of PMOS transistors in the *s208* circuit, both before and after the application of the proposed method. The horizontal axis represents $SP_0$ values, while the vertical axis shows the number of nodes corresponding to each $SP_0$. The results demonstrate a significant reduction in the number of high-$SP_0$ nodes after applying the method, effectively mitigating NBTI-induced degradation. Notably, no nodes with $SP_0 > 0.8$ remain in the modified circuit. It is worth noting that the number of PMOS transistors differs before and after applying the proposed method.

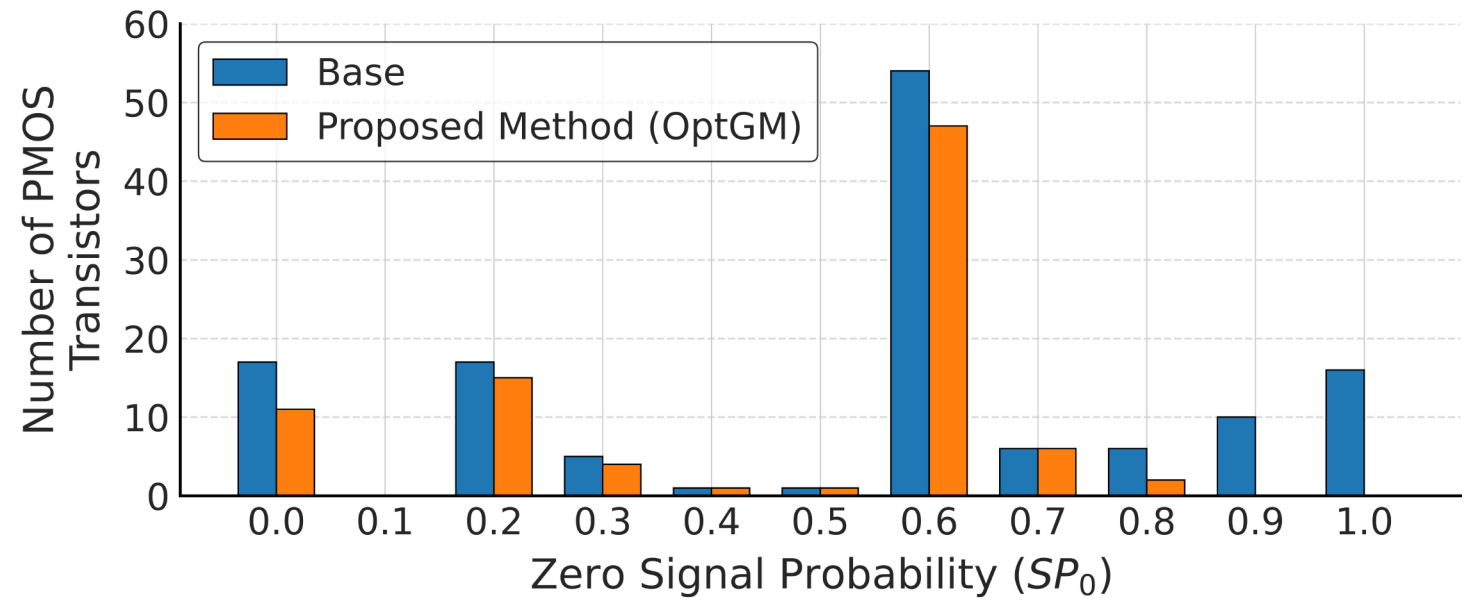

Fig. 8. $SP_0$ distribution of PMOS transistors in the ***s208*** circuit before and after applying the proposed method.

Table 2 reports the critical path delays of various benchmark circuits before and after NBTI stress, both for the base circuits and for those improved using OptGM. The second and fourth columns show delays before stress, while the third and sixth columns show delays after 10 years of NBTI aging. The optimal $SP_0$ threshold value used for each circuit is listed in the fifth column. Columns seven and eight indicate the percentage of delay improvement achieved by the OptGM method before and after NBTI stress, respectively. On average, the proposed method improves pre-stress delay by 7.89% and post-stress delay by 24.03%. Statistical analysis with a 95% confidence level indicates that OptGM achieves an NBTI-induced delay degradation improvement within the range of 16.73% to 31.01% for similar circuits using the same technology model.

Table 2. Delay results for the base and OptGM-enhanced circuits before and after NBTI stress

| Circuit | Base circuit | | Proposed method (OptGM) | | | OptGM method improvement | |
|---|---|---|---|---|---|---|---|
| | Delay (*ps*) before stress | Delay (*ps*) after stress | Delay (*ps*) before stress | Optimal threshold value | Delay (*ps*) after stress | Delay (%) before stress | Delay (%) after stress |
| *s27* | 20.2 | 40.5 | 20.3 | 0.75 | 38.6 | -0.50 | 4.69 |
| *s208* | 176.0 | 287 | 166 | 0.75 | 247 | 5.68 | 13.94 |
| *s298* | 263.0 | 449 | 260 | 0.65 | 352 | 1.14 | 21.60 |
| *s400* | 15.9 | 27.2 | 15.8 | 0.65 | 20.2 | 0.63 | 25.73 |
| *c432* | 189.2 | 352.2 | 159.6 | 0.5 | 203 | 15.64 | 42.36 |
| *c1355* | 169.0 | 284.0 | 161.0 | 0.65 | 229.5 | 4.73 | 19.21 |
| *c2670* | 227.5 | 402.0 | 208.1 | 0.85 | 320.1 | 8.49 | 20.37 |
| *c1908* | 229.8 | 392.3 | 210.8 | 0.85 | 320.7 | 8.24 | 18.25 |
| *c3540* | 290.2 | 487.7 | 252.7 | 0.75 | 327.6 | 12.93 | 32.81 |
| *c5315* | 221.2 | 391.0 | 185.8 | 0.65 | 285.7 | 16.02 | 26.94 |
| *c6288* | 928.7 | 1514.4 | 749.5 | 0.5 | 959.8 | 19.29 | 36.62 |
| *b14* | 198.5 | 341.2 | 186.3 | 0.75 | 251.5 | 6.14 | 26.3 |
| *b15* | 215.7 | 368.9 | 203.2 | 0.65 | 280.1 | 5.79 | 24.1 |
| *b17* | 312.8 | 531.6 | 289.4 | 0.85 | 415.7 | 7.48 | 21.8 |
| *b20* | 247.3 | 422.1 | 230.8 | 0.75 | 313.6 | 6.67 | 25.7 |
| **Average** | | | | | | **7.89** | **24.03** |

While OptGM achieves a 24.03% reduction in NBTI delay, co-optimization of multiple aging mechanisms represents a valuable research direction. For example, one practical route is dual-threshold selection—introducing separate thresholds for $SP_0$ (NBTI) and $SP_1$ (PBTI) and guiding choices with a modified metric such as $PPC' = 1/(Delay_{NBTI} \times Delay_{PBTI} \times Area)$—and complementing this with HCI-aware gate sizing, i.e., optimizing transistor widths inside merged complex cells to limit drain-to-source voltage and reduce hot-carrier stress.

Table 3 presents the transistor count, number of NBTI-critical PMOS transistors, and area overhead before and after applying OptGM. On average, OptGM eliminates 89.3% of NBTI-critical transistors (i.e., PMOS transistors connected to critical nodes). It is worth mentioning that, while the proposed method primarily targets the elimination of PMOS transistors that are frequently on, the merging of gates also results in the removal of associated NMOS transistors, which are typically in the off state. Consequently, the likelihood of increased NMOS stress—and thus PBTI degradation—is minimal, or may even be slightly reduced. Furthermore, most circuits utilize fewer total transistors after modification, due to the consolidation of logic into complex gates. However, in cases where sensitizer gates have multiple fan-out branches, duplication of gates is required for each sensitive gate, potentially increasing the area, which results in an average 1.96% increase in area. For example, in the *s27* circuit, two sensitizer gates with dual fan-outs lead to duplication, increasing the area. Conversely, in many other circuits, the area reduction from gate merging exceeds the overhead from duplication.

Table 3. Area results of base circuits and those modified with OptGM

| Circuit | Base circuit | | | Proposed method (OptGM) | | | OptGM method improvement | | Area overhead (%) |
|---|---|---|---|---|---|---|---|---|---|
| | Total transistors (#) | Critical transistors (#) | Sum of transistors' width (*mm*) | Total transistors (#) | Critical transistors (#) | Sum of transistors' width (*mm*) | Reduction of total transistors (%) | Reduction of critical transistors (%) | |
| *s27* | 42 | 3 | 3.98 | 48 | 0 | 5.60 | -14.3 | 100.0 | 40.7 |
| *s208* | 478 | 32 | 64.9 | 448 | 2 | 62.28 | 6.3 | 93.8 | -4.0 |
| *s298* | 782 | 66 | 39.54 | 693 | 8 | 37.83 | 11.4 | 87.9 | -4.3 |
| *s400* | 850 | 26 | 95.28 | 795 | 0 | 76.89 | 6.5 | 100.0 | 1.3 |
| *c432* | 832 | 76 | 139.58 | 736 | 5 | 98.8 | 11.5 | 93.4 | -19.2 |
| *c1355* | 2356 | 104 | 201.28 | 2116 | 10 | 212.53 | 10.1 | 90.4 | 5.6 |
| *c2670* | 1526 | 218 | 789.12 | 1366 | 54 | 729.34 | 10.5 | 75.2 | 7.57 |
| *c1908* | 5082 | 199 | 495.49 | 4584 | 32 | 486.3 | 9.8 | 83.91 | 1.86 |
| *c3540* | 2167 | 532 | 1153.93 | 1977 | 88 | 1116.43 | 8.77 | 83.45 | 3.35 |
| *c5315* | 3025 | 754 | 1670.94 | 3014 | 105 | 2005.13 | 0.36 | 86.07 | -0.2 |
| *c6288* | 2672 | 890 | 1591.2 | 2396 | 96 | 1788.50 | 10.32 | 89.21 | -12.4 |
| *b14* | 1836 | 167 | 312.45 | 1673 | 14 | 318.98 | 8.88 | 91.4 | 2.1 |
| *b15* | 1914 | 183 | 335.67 | 1751 | 21 | 342.05 | 8.51 | 88.7 | 1.9 |
| *b17* | 6498 | 563 | 1127.83 | 5942 | 78 | 1159.41 | 8.56 | 86.2 | 2.8 |
| *b20* | 4071 | 392 | 702.18 | 3738 | 40 | 718.33 | 8.18 | 89.9 | 2.3 |
| **Average** | | | | | | | **7.03** | **89.30** | **1.96** |

Table 4. Average $SP_0$ of benchmark circuits before and after applying the proposed method

| Circuit | Average $SP_0$ of base circuit | Average $SP_0$ of modified circuit | Average $SP_0$ reduction (%) |
|---|---|---|---|
| *s27* | 0.56 | 0.53 | 5.36 |
| *s208* | 0.47 | 0.34 | 27.66 |
| *s298* | 0.53 | 0.38 | 28.30 |
| *s400* | 0.49 | 0.28 | 42.86 |
| *c432* | 0.48 | 0.30 | 37.50 |
| *c1355* | 0.46 | 0.32 | 35.43 |
| *c2670* | 0.41 | 0.34 | 15.93 |
| *c1908* | 0.58 | 0.46 | 20.16 |
| *c3540* | 0.69 | 0.39 | 33.47 |
| *c5315* | 0.45 | 0.34 | 28.88 |
| *c6288* | 0.49 | 0.29 | 40.82 |
| *b14* | 0.52 | 0.36 | 30.77 |
| *b15* | 0.54 | 0.38 | 29.63 |
| *b17* | 0.61 | 0.43 | 29.51 |
| *b20* | 0.57 | 0.39 | 31.58 |
| **Average** | | | **29.19** |

Table 4 reports the average $SP_0$ values before and after applying OptGM. The method achieves a mean reduction in $SP_0$ of 29.19%, effectively reducing the NBTI stress time of transistors. Since NBTI-induced threshold voltage shifts are strongly time-dependent, reducing stress duration leads to substantial reliability improvement.

Fig. 9 shows the percentage of NBTI-induced delay degradation reduction for different durations of NBTI stress—5, 10, 30, and 50 years—under a fixed threshold value of 0.75. The base and modified circuits have been simulated for different years under NBTI stress, and the percentage of delay improvement has been reported. The results show that the improvement percentage increases when the stress time increases. Therefore, the proposed method increases performance in longer stress time efficiently. It should be noted that in the simulation results of Fig. 9, for stress times beyond 10 years, NBTI saturation effects were also taken into account.

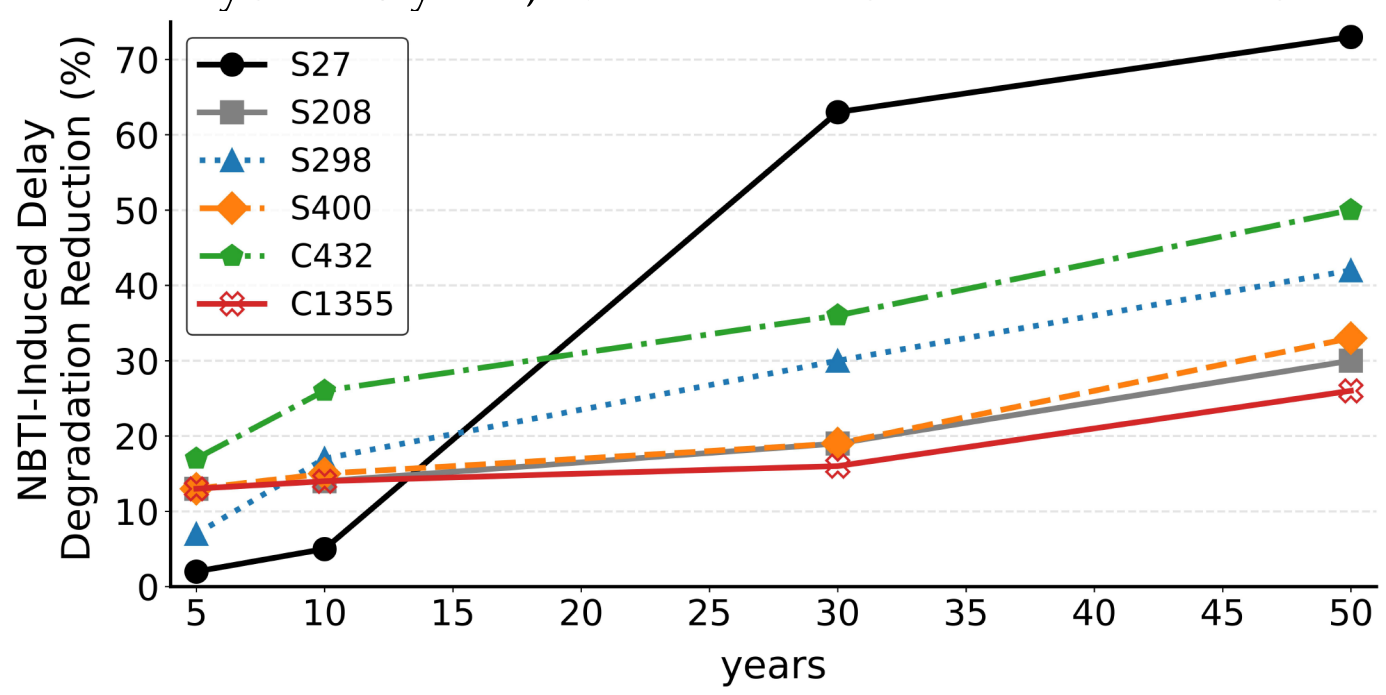

Fig. 9. The percentage of NBTI-induced delay degradation reduction under 5, 10, 30, and 50 years of NBTI stress on benchmark circuits (OptGM modified vs. base circuit).

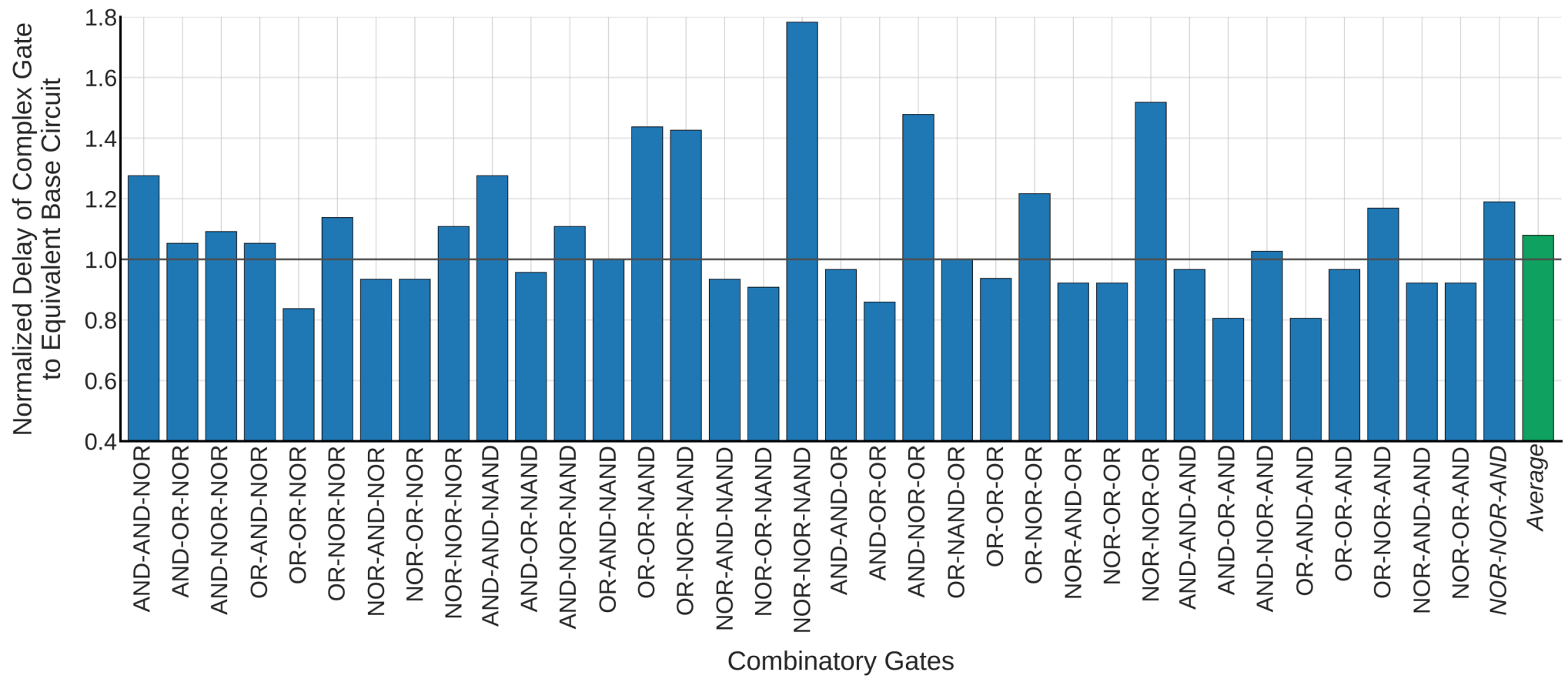

(a)

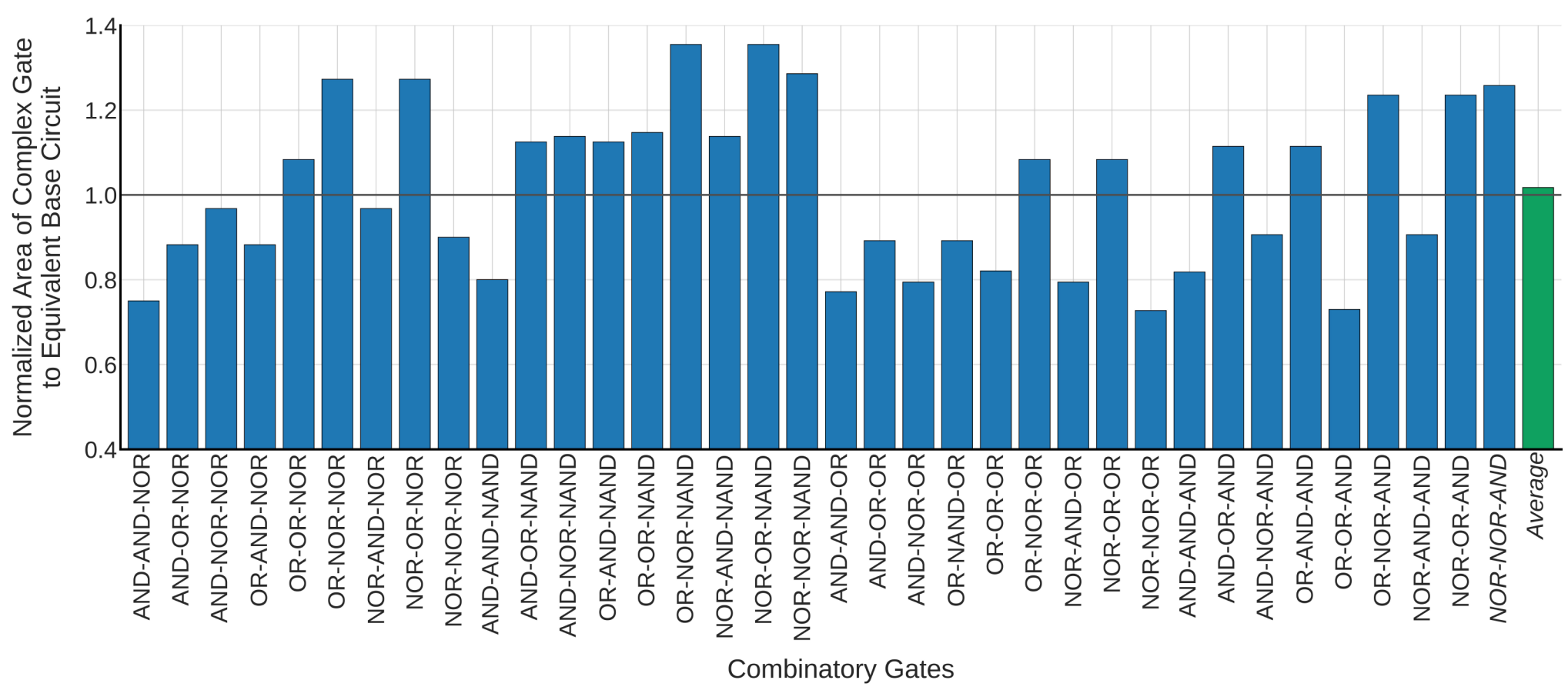

(b)

Fig. 10. Comparison of original circuits and their equivalent complex gates in terms of a) delay and b) area.

Fig. 10(a) and Fig. 10(b) present a comparison of delay and area, respectively, for original circuits and their equivalent complex gates. The vertical axis shows the ratio of the complex gate metric to the equivalent metric in its base circuit. Based on the results, the average delay of complex gates is approximately 7% higher than that of their equivalent primary-gate implementations. However, Fig. 10(b) shows that the average complex gate area is nearly equal to that of its equivalent circuit. Note that when *n* fan-outs exist for a sensitizer gate, *n* identical copies are generated—contributing to area growth in some cases.

OptGM induces two competing effects in the power trade-off: fewer transistors, which reduces leakage power, versus potential increases in internal capacitance, which raise dynamic power. On average, the results show an 8.3% reduction in leakage power and a 1.8% increase in dynamic power, yielding a net power benefit in most cases. Nonetheless, unfavorable scenarios do exist; in some circuits, the complex gate exhibits higher internal capacitance than the replaced network, resulting in a dynamic-power penalty (with the worst observed case being 12.3% for c5315 at the aggressive threshold of 0.5). In this regard, a practical mitigation strategy can be adopted. For example, a power-aware variant of the PPC metric can enforce power neutrality while still achieving significant NBTI reduction, or the power constraint can be enabled for power-critical designs while the original PPC is used for performance-critical flows.

In order to validate the simulations with a practical implementation, the ASIC hardware design flow was followed, and experimental ASIC-prototype parameter results were obtained. In this regard, three benchmark circuits were synthesized with Synopsys Design Compiler targeting a 500MHz clock using the 45*nm* standard-cell library under moderate effort constraints. The synthesized netlists were processed by Algorithm 1 to identify and instantiate merges; custom complex cells were characterized in HSPICE and converted to Liberty format, producing an extended library of 18 complex-gate variants. Place-and-route was performed with Cadence Innovus using identical constraints for the baseline and OptGM designs, parasitics were extracted with Cadence Quantus QRC, and final post-layout static timing analysis (STA) was conducted in Synopsys PrimeTime. Measured quantities included setup slack, critical-path delay, and actual placed area, and power was measured and reported in Table 5.

Table 5. Comparison of hardware implementation parameters of three base circuits and those modified with OptGM

| Circuit | Version | Post-P&R Frequency (MHz) | Critical Path Delay (ns) | Slack (ns) | Area (um$^2$) | Power (mW) |
|---|---|---|---|---|---|---|
| *s208* | Base | 476 | 2.10 | -0.10 | 1,847 | 1.83 |
| | OptGM | 501 | 2.00 | +0.01 | 1,881 | 1.79 |
| | Change | +5.25% | -4.76% | +0.11 | +1.84% | -2.19% |
| *c432* | Base | 441 | 2.27 | -0.27 | 3,024 | 2.91 |
| | OptGM | 489 | 2.04 | -0.04 | 3,079 | 2.85 |
| | Change | +10.88% | -10.13% | +0.23 | +1.82% | -2.06% |
| *c1355* | Base | 463 | 2.16 | -0.16 | 6,138 | 5.47 |
| | OptGM | 494 | 2.02 | -0.02 | 6,247 | 5.31 |
| | Change | +6.69% | -6.48% | +0.14 | +1.78% | -2.92% |

The findings validate the simulation results and show four points. First, OptGM yields a pre-aging frequency improvement after P&R, confirming that merged gates do not introduce a layout-level delay penalty. Second, the observed post-layout area overhead matches the transistor-count-based prediction, validating our area estimation approach. Third, measured power after layout shows a modest improvement; the small discrepancy can be attributed to routing capacitances not modeled in the HSPICE cell-level characterization. Fourth, these three OptGM modified circuits achieved timing closure (positive or near-zero slack) with no introduced timing violations. In conclusion, the results indicate that the benefits of OptGM are likely to translate to real hardware.

**Comparison with Related Work:** The proposed OptGM method significantly reduces NBTI-induced degradation by 24.03% and effectively eliminates 89.3% of NBTI-critical transistors, achieving a 12.8% improvement in PPC compared to the previous gate merging approach [14], as shown in Table 6. This enhancement results from OptGM's adaptive threshold selection and systematic gate merging algorithms. Table 6 compares PPC values for gate merging with fixed and variable thresholds, highlighting the importance of using an adaptive threshold over the method in [14].

Table 6. Comparison of PPC for gate merging methods with fixed and variable threshold values

| Circuit | PPC value of method [14] | Threshold value of OptGM method | PPC of OptGM method | PPC improvement (%) |
|---|---|---|---|---|
| *S27* | 1.77 | 0.75 | 1.77 | 0.00 |
| *S280* | 1.12 | 0.75 | 1.12 | 0.00 |
| *S298* | 1.21 | 0.65 | 1.22 | 0.81 |
| *S400* | 1.00 | 0.65 | 1.36 | 36.25 |
| *C432* | 1.11 | 0.50 | 1.46 | 31.17 |
| *C1355* | 1.20 | 0.65 | 1.31 | 8.93 |
| **Average** | | | | **12.80** |

Besides method [14], various methods have been proposed to address NBTI. We conducted a comparative analysis against five NBTI mitigation methods spanning different categories (INC, IVC, and adaptive techniques), and the results are summarized in Table 7.

Table 7. Comparison of OptGM with related works in terms of different parameters

| Method | Category | Technology | NBTI Delay Reduction | Area Overhead | Power Overhead | Benchmarks | Limitations |
|---|---|---|---|---|---|---|---|
| *Lin et al. [15]* | INC | 90*nm* | 32.36% | 18.4% | 15.2% | 5 ISCAS circuits | Transmission gates increase both area and power; high overhead limits scalability |
| *Wang et al. [17]* | INC | 65*nm* | ~18.5% | 11.8% | N/A | 8 ISCAS circuits | Library-constrained gate replacement; cannot handle all critical nodes; timing impact not optimized |
| *Bild et al. [31]* | INC | 45*nm* | 1.21% | 7.9% | -5.1% | 6 ISCAS circuits | Minimal NBTI reduction; requires custom standard cell redesign; structural modifications complex |
| *Kumar et al. [36]* | Adaptive | 65*nm* | ~25% | 0% | 9.8% | Analytical | Forward body biasing exponentially increases leakage; runtime energy penalty; requires substrate contacts |
| *Kiamehr et al. [13]* | IVC | 45*nm* | 14.2% | 0% | 3.4% | 7 ISCAS circuits | Limited to shallow nodes controllable by inputs; deep logic levels unaffected; effectiveness decreases with circuit depth |
| *Ghane et al. [14]* | INC | 45*nm* | 20.12% | 2.5% | N/A | 6 ISCAS circuits | Fixed threshold (0.5); no optimization framework |
| *OptGM (ThisWork)* | INC | 45*nm* | 24.03% | 1.96% | -3.2% | 15 ISCAS circuits | Primarily targets combinational logic; flip-flop aging requires complementary techniques |

In summary, OptGM achieves a favorable trade-off between effectiveness and implementation overhead across all comparisons. Compared with Lin et al. [15], which attains the highest NBTI reduction via transmission-gate insertion, OptGM attains approximately 74% of that reduction while incurring only a fraction of the cost: Lin's method requires roughly 18.4% area overhead and a 15.2% power penalty, whereas OptGM achieves comparable mitigation with 1.96% area overhead and an overall power reduction (−3.2%). When normalized by our PPC metric (1/(Delay×Area)), OptGM's PPC is 3.8× that of Lin [15], highlighting a substantially better power–performance–area trade-off.

Relative to Wang et al. [17], which relies on library-based gate replacement, OptGM is both more effective and more general. Wang's approach is limited by available standard cells—less than half of NBTI-critical nodes in our experiments had suitable replacements—whereas OptGM performs on-demand gate merging to eliminate critical nodes comprehensively. Consequently, OptGM yields higher NBTI reduction while producing roughly lower area overhead. Bild et al. [31], which requires extensive modifications to pull-up/pull-down networks across the standard-cell library and achieves modest leakage reduction, delivers negligible NBTI mitigation based on our evaluation. By contrast, OptGM attains an order-of-magnitude larger NBTI benefit.

Adaptive substrate-bias techniques, such as Kumar et al. [36], avoid area overhead but trade off increased leakage and energy consumption; in our analysis, a representative adaptive scheme incurs a 9.8% power penalty due to leakage growth. Over a 10-year horizon, accounting for both dynamic and leakage energy, OptGM yields approximately high energy savings relative to the adaptive approach, because OptGM reduces the root causes of aging rather than compensating for them at the expense of power. Finally, software-only IVC approaches exemplified by Kiamehr et al. [13] impose zero area overhead but are fundamentally depth-limited: their effectiveness decays rapidly with logic depth and, for deep circuits such as b17 (depth = 23), IVC achieves only 6.8% reduction compared with OptGM's 21.8% on the same design. On average across our benchmarks, OptGM provides roughly greater NBTI reduction than IVC by directly controlling internal nodes that are unreachable by input-vector manipulation.

## 6- Conclusion

As technology scales down, NBTI has become a major reliability concern in modern CMOS circuits due to its impact on PMOS threshold voltage and circuit delay. This paper proposed OptGM, an optimized gate merging method that effectively mitigates NBTI-induced degradation by eliminating NBTI-critical internal nodes through SP0-based analysis and logic gate merging. Unlike prior methods, OptGM adaptively selects the optimal threshold per circuit using a performance-per-cost metric, enabling a better trade-off between NBTI mitigation and overhead. Simulation results on standard benchmarks demonstrated that OptGM reduces NBTI-induced delay degradation by 24.03%, removes 89.3% of NBTI-critical transistors, and improves PPC by 12.8%, with minimal area overhead—highlighting its effectiveness. As future work, the development of advanced optimization metrics incorporating multiple design parameters with tunable priorities, along with the integration into logic synthesis tools, can be explored to support fully automated NBTI-aware ASIC design flows. In addition, a co-optimization gate merging method addressing both NBTI and PBTI can be introduced, which becomes particularly important in advanced technology nodes.

## Appendix

**Formal analysis of the correctness and completeness of the proposed algorithms.**

**Correctness (logic equivalence preservation)**

Statement: For any gate merging operation performed by Algorithm 3, the output logic function of the resulting complex gate is equivalent to the original cascade of sensitizer and sensitive gates.

Proof:
Let

- $G_{sens}$ = sensitizer gate with output node n and logic function $f(x_1, x_2, ..., x_k)$
- $G_{sent}$ = sensitive gate with inputs including n and logic function $g(n, y_1, y_2, ..., y_m)$
- Original cascade output: $O_{original} = g(f(x_1, ..., x_k), y_1, ..., y_m)$

Algorithm 3 (line 13) generates a complex gate $G_{complex}$ by:

- Substituting $n = f(x_1, ..., x_k)$ into $g(n, y_1, ..., y_m)$
- Applying Boolean algebra simplification using standard complex gate synthesis rules
- The output of $G_{complex}$ is: $O_{complex} = g(f(x_1, ..., x_k), y_1, ..., y_m)$

By construction: $O_{original} \equiv O_{complex}$ (Boolean equivalence)

Formal verification was performed using Cadence Conformal LEC (Logic Equivalence Checking) for all merged gates in all benchmark circuits, and zero equivalence violations were detected.

**Completeness (Critical node identification)**

Statement: Algorithm 2 identifies all internal nodes with $SP_0 \geq$ Threshold in a given circuit netlist.

Proof by Induction on circuit depth:

Base case (depth L = 0, primary inputs): Line 5 assigns $SP_0 = 0.5$ to all primary inputs. All depth-0 nodes are processed.

Inductive hypothesis: Assume all nodes at depth ≤ k have correct $SP_0$ values computed.

Inductive step (depth k+1): Lines 6-10 iterate through all gates at depth k+1 (topologically sorted, line 1). For each gate g at depth k+1:

- All inputs of g are at depth ≤ k (by definition of topological order)

- By the inductive hypothesis, input $SP_0$ values are correct
- Line 8 computes output $SP_0$ using Table 1 formulas (proven correct in Boolean probability theory)
- Therefore, $SP_0$(output of g) is correct

Lines 12-16 iterate through all nodes N = {$n_1$, $n_2$, ..., $n_m$} and classify each as critical or non-critical based on the threshold. No node is skipped (foreach loop guarantees completeness).

In conclusion, by induction, all nodes at all depths have correct $SP_0$ values, and all critical nodes are identified.